\documentclass[3p,times]{elsarticle}

\usepackage{ecrc}

\volume{00}

\firstpage{1}

\journalname{International Journal of Non-Linear Mechanics}

\runauth{B. Kaviya et al.}

\jid{CNSNS}

\jnltitlelogo{Int J Nonlin Mech}

\usepackage{amssymb}
\usepackage{hyperref}
\usepackage{amsmath}

\begin{document}
\begin{frontmatter}
		
	
		
\dochead{}
		
\title{Influence of dissipation on extreme oscillations of a forced anharmonic oscillator}
		
		
\author[label1]{B. Kaviya}
\author[label1]{R. Suresh \corref{cor1}}
\ead{sureshphy.logo@gmail.com}
\author[label1]{V. K. Chandrasekar}
\ead{chandru25nld@gmail.com}
\author[label2]{B. Balachandran}
\ead{balab@umd.edu}
\address[label1]{Centre for Nonlinear Science and Engineering, School of Electrical and Electronics Engineering, SASTRA Deemed University, Thanjavur 613 401, India}
\address[label2]{Department of Mechanical Engineering, University of Maryland, College Park, MD 20742, USA}
\cortext[cor1]{Corresponding author}
		
\begin{abstract}
Dynamics of a periodically forced anharmonic oscillator with cubic nonlinearity, linear damping, and nonlinear damping, is studied. To begin with, the authors examine the dynamics of an anharmonic oscillator with the preservation of parity symmetry. Due to this symmetric nature, the system has two neutrally stable elliptic equilibrium points in positive and negative potential-wells. Hence, the unforced system can exhibit both single-well and double-well periodic oscillations depending on the initial conditions. Next, the authors include position-dependent damping in the form of nonlinear damping ($x\dot{x}$) into the system.  Then, the parity symmetry of the system is broken instantly and the stability of the two elliptic points is altered to result in stable focus and unstable focus in the positive and negative potential-wells, respectively. Consequently, the system is dual-natured and is either non-dissipative or dissipative, depending on location in the phase space. The total energy of the system is used to explain this dual nature of the system. Furthermore, when one includes a periodic external forcing with suitable parameter values into the nonlinearly damped anharmonic oscillator system and starts to increase the damping strength, the parity symmetry of the system is not broken right away, but it occurs after the damping reaches a threshold value. As a result, the system undergoes a transition from double-well chaotic oscillations to single-well chaos mediated through a type of mixed-mode oscillations called extreme events (EEs) in which the small-amplitude single-well chaotic oscillations are interrupted by rare and recurrent large-amplitude (double-well) chaotic bursts. Furthermore, it is found that the large-amplitude oscillations developed in the system are completely eliminated if one incorporates linear damping into the system.  Hence, it is believed that a novel means has been identified for controlling the EEs that occur in forced anharmonic oscillator system with nonlinear damping. The numerically calculated results are in good agreement with the theoretically obtained results on the basis of Melnikov's function. Further, it is demonstrated that when one includes linear damping into the system, this system has a dissipative nature throughout the entire phase space of the system.  This is believed to be the key to the elimination of EEs.
\end{abstract}
		
\begin{keyword}
Anharmonic oscillator \sep Position dependent damping \sep Nonlinear damping \sep mixed-mode oscillations \sep Extreme events \sep

\end{keyword}
		
\end{frontmatter}
	

\section{Introduction}
\label{sec1}
Forced and damped nonlinear oscillators have been considered as paradigms for mimicking the dynamics of various physical and engineering systems such as  Josephson junctions, electrical circuits, optical systems, macromechanical and microelectromechanical oscillators, and so on \cite{guckenheimer1983, kovacic2011, lakshmanan1996, ji2002, li2006}. \emph {Damping}, which is used to model loss of energy due to friction and viscous forces, is ubiquitous in many mechanical systems and this characteristic influences the performance of the oscillators in different ways. Most oscillatory systems are subject to different damping combinations and each one of them has different effects on the considered dynamical systems. In general, it is common to use a linear damping model for describing damping or dissipation experienced by a system. However, in many oscillatory systems such as microelectromechanial and nanoelectromechanical oscillators, nonlinear damping is found to play a significant role. For example, in nanoelectromechanical systems made from carbon nanotubes and graphene, damping is found to strongly depend on the amplitude of motion, and the damping force is nonlinear in nature \cite{eichler2011}.  These systems are being used for mass and force sensing applications \cite{ekinci2004, papariello2016}. Researchers have exploited the nonlinear nature of damping in these systems to improve the figures of merit for both nanotube and graphene resonators.  In reference \cite{akerman2010}, the ion steady-state motion is well described by the Duffing oscillator model with an additional nonlinear damping term.  Both the linear damping and nonlinear damping can be tuned with the laser-cooling parameters helping one to investigate the mechanical noise squeezing in laser cooling. Recently, the influence of nonlinear damping on the motion of a nanobeam resonator was studied and it was found that nonlinear damping can have a significant impact on the dynamics of micromechanical systems \cite{zaitsev2012}. In fluid mechanics, linearly forced isotropic turbulence can be described by an anharmonic oscillator model with nonlinear damping \cite{ran2009}.  From a dynamics viewpoint, it has been shown that nonlinear damping can be used to suppress chaos in oscillatory systems \cite{siewe2009, miwadinou2015, ravindra1995}. The stability of responses of nonlinearly damped, hard and soft Duffing oscillators have also been analyzed \cite{ravindra1994, ravindra1994a}. In addition, the effect of nonlinear damping in forced Duffing and other types of nonlinear oscillators has been extensively studied \cite{bikdash1994, nayfeh1995, almog2007, baltanas2001, sanjuan1999, jing2009}.
	
Nonlinear damping plays a significant role in the dynamics of systems driven by a direct external periodic forcing or a parametric excitation \cite{kovacic2011, leuch2016, lifshitz2008,patidar2016}. Specifically, the development of mixed-mode oscillations and extreme events (EEs) have been recently reported in systems influenced by nonlinear damping \cite{kingston2017, kingston2017a, kingston2018, suresh2018, suresh2019}. The rare and recurrent occurrence of large-amplitude events in system variables with heavy tails in the probability distribution is a signature of EEs. Examples of EEs that occur in natural and engineering systems include rogue waves in optical systems and oceans, epidemics, large-scale power black-outs in electrical power grids, harmful algal blooms in marine ecosystems, jamming in computer and transportation networks, stock market crashes, and epileptic seizures \cite{dysthe2008, chabalko2014, solli2007, chen2015, lehnertz2008, bialonski2016, dobson2007,chowdhury2019,ray2019}. Similar statistical behaviors of the appearance of sudden changes in the system variables have been noticed in many dynamical systems governed by nonlinear equations with nonlinear damping \cite{kingston2017a, suresh2018, suresh2019}. However, an understanding of the occurrence of EEs in such systems is still being developed and the significance of nonlinear damping for the development of EEs has not received careful attention. Furthermore, an understanding of the mechanism that triggers EEs in dynamical systems is crucial for developing strategies to control such events. Although this is out of reach in natural systems, it may certainly be possible in several engineering systems, such as power grid networks, mechanical systems, optical systems, and so on. In these systems, one can design control techniques to avoid the emergence of EEs. In line with this, control of EEs in dynamical systems has been recently investigated \cite{suresh2018, chen2015, chen2014, cavalcante2013, galuzio2014}. However, the studies carried out in this direction are quite limited and a systematic study on control of EEs is still in the early stages of research. Motivated by the above, in this paper, the authors investigate the dynamics of an anharmonic oscillator with cubic nonlinearity in the presence of linear damping, nonlinear damping, and periodic external forcing. A primary objective of this paper is to establish an understanding of the role played by the nonlinear damping in the development of EEs and strategies to control such events.
	
First, the authors study the dynamics of the undamped, anharmonic oscillator, in which the parity ($\mathcal{P}$)-symmetry is preserved. Due to this symmetric nature, the system has two neutrally stable elliptic equilibrium points in both positive and negative potential-wells. Therefore, the system has a conservative nature in the entire phase space and can exhibit single-well periodic oscillations if the trajectories are started near one of the equilibrium points or double-well periodic oscillations when the initial conditions are chosen away from these equilibrium points. It is shown that the system motions are single-well periodic oscillations when the initial conditions are chosen from the region where the total energy is negatively valued. On the contrary, the system motions are in the form of double-well periodic oscillations if the initial conditions are chosen in the region where the total energy is positively valued.
	
Next, the authors add a position-dependent damping or nonlinear damping term of the form $\alpha x\dot{x}$ into the anharmonic oscillator equation and investigate the system dynamics with respect to the nonlinear damping parameter. Due to the inclusion of nonlinear damping term, the symmetry of the system is broken instantly and the system has a parity and time-reversal ($\mathcal{PT}$) - symmetry; this alters the stability of the equilibrium points. For the positive values of $\alpha$, the equilibrium point in the negative potential-well becomes a source and repels nearby trajectories. The repelled trajectories are attracted by the fixed point in the right potential-well that acts as a sink. Therefore, the system has a dissipative nature in some regions of the phase space in which the trajectories are damped and attracted to the right potential-well. At the same time, the system has non-dissipative dynamics in other areas of phase space where the trajectories are in the form of periodic oscillations. The system has either a dissipative or a non-dissipative nature, depending on the location in the phase space. The underlying mechanism is explained in terms of the total energy of the system.
	
Furthermore, when one considers an external periodic forcing of the anharmonic oscillator, in the absence of nonlinear damping, the system preserves symmetry and the system motions are manifested as double-well chaotic oscillations for certain values of amplitude and frequency of the external forcing. As earlier mentioned, the inclusion of nonlinear damping makes the system asymmetric and the unstable focus in the left potential-well does not attract system trajectories. Therefore, the number of trajectories travelling into the left potential-well is gradually reduced as a function of the nonlinear damping strength. The system exhibits large-amplitude oscillations that are alternated with small-amplitude oscillations.  These oscillations are named as bursting-like oscillations (BOs). Specifically, for a certain range of the nonlinear damping parameter, the large-amplitude (double-well) oscillations occur sporadically and recurrently with a highly unpredictable nature. These rarely occurring large-amplitude oscillations are characterized as EEs. To differentiate EEs from other dynamical states, the threshold $H_s = \langle P_{n}\rangle + 8 \sigma$ has been numerically estimated \cite{kingston2017a}. Here, $\langle P_{n}\rangle$ is the time-averaged peak value of one of the system variables and $\sigma$ stands for the mean standard deviation. In other words, the threshold height is equal to the time-averaged mean value of the peak plus eight times the standard deviation derived for a long run with the iterations of $2\times10^{9}$ time units (after leaving out transients). During the occurrence of EEs the large-amplitude oscillations occur occasionally. Therefore, the peaks are larger than the threshold $H_s$. By contrast, for the other dynamical states, the average peak value ($\langle P_{n}\rangle$) is quite high. Hence, $H_s$ becomes higher than the large peaks. Finally, the system exhibits single-well bounded chaotic oscillations when one increases the damping strength above the threshold value. In a nutshell, the authors have found that by including the nonlinear damping term into the forced anharmonic oscillator system, the $\mathcal{P}$--symmetry is not broken instantaneously. But this happens only when the damping parameter is taken beyond a threshold value. As a result, the system undergoes a transition from double-well chaotic oscillations to single-well chaos intervened by EEs, with respect to variation in the nonlinear damping parameter.
	
In accordance with the goal of suppressing large-amplitude oscillations and to identify means to control EEs, a linear damping term is included in the forced anharmonic oscillator along with nonlinear damping and the authors examine the responses of the resulting dynamical system. Interestingly, it is found that the large-amplitude oscillations are completely eradicated from the system dynamics and only single-well small-amplitude oscillations are feasible.  The authors also show that the elimination of EEs occurs through two different dynamical routes as a function of the forcing frequency and the strength of nonlinear damping. One is a transition from EEs to periodic oscillations, and another is a transition from BOs to single-well oscillations intervened by EEs. The authors' findings are supported by both numerical and theoretical results, which include bifurcation diagram plots and Melnikov function estimates \cite{han2012}. It is remarked that the theoretically determined results are in good agreement with the numerically obtained results. In addition, the mechanism for the elimination of EEs is examined and the authors have found that the inclusion of linear damping destroys the non-dissipative nature of the system, which attains a dissipative nature throughout the entire phase space of the system in the absence of external forcing. Consequently, the trajectories initiated anywhere in the phase space follow a decaying solution, which is believed to be a key for the suppression of large-amplitude oscillations.
	
The remainder of this paper has been organized as follows: In Section ~\ref{sec2}, the authors study the dynamics of an anharmonic oscillator with and without nonlinear damping and demonstrate the non-trivial property of the coexistence of dissipative and conservative nature of the nonlinearly damped anharmonic oscillator. Section \ref{sec3} is devoted to the study of the forced anharmonic oscillator with nonlinear damping in which the transition from double-well to single-well chaotic oscillations mediated by EEs and the response changes observed with respect to damping strength variation are presented. Control of EEs through the inclusion of linear damping into the system is illustrated in Section ~\ref{sec4}.  Following that, in the next section, the mechanism underlying suppression of EEs is examined. Finally, in Section ~\ref{sec5}, the authors collect together their conclusions.
\section{Dynamics of an unforced, anharmonic oscillator without and with nonlinear damping}
\label{sec2}
\begin{figure}
\centering \includegraphics[width=1.0\columnwidth]{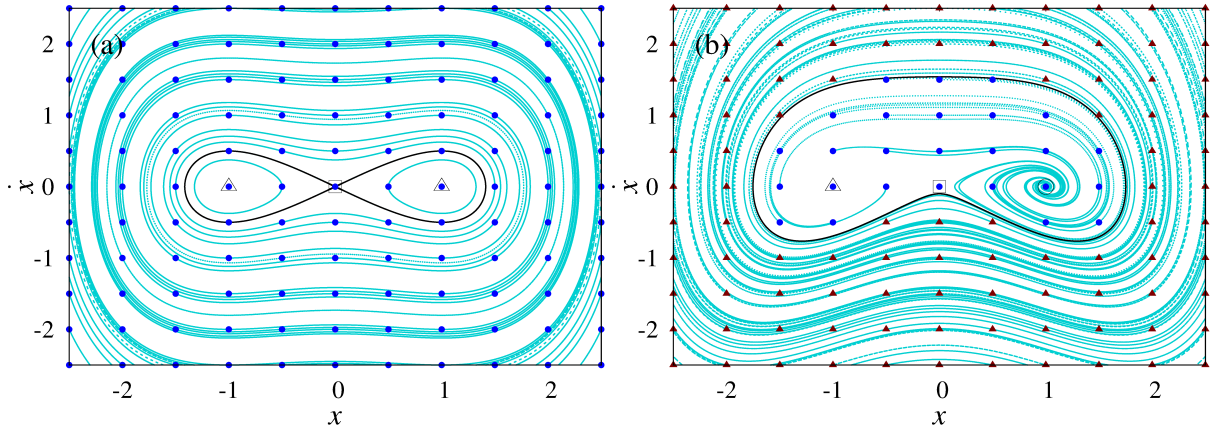}
\caption{(a) Phase portrait of undamped and unforced, anharmonic oscillator (\ref{eqn1}) with $\gamma=0.5$ and $\beta=0.5$. Open triangles are used at ($\pm1,0$) to denote elliptic equilibrium points and a square is used at (0,0) to denote a saddle point. The chosen initial conditions are indicated by filled circles and the trajectories that are initiated from them are depicted with aqua (light gray) lines. The closed loop, which is shown in black, is a homoclinic orbit. (b) Phase portrait of unforced, anharmonic oscillator with nonlinear damping (\ref{eqn3}) and $\alpha=0.45$. The filled circles within the homoclinic orbit represent initial conditions, from which the trajectories that follow have a dissipative nature. The filled triangles represent initial conditions, from which the trajectories that follow have a non-dissipative nature. The corresponding motions are periodic oscillations.}
\label{fig1}
\end{figure}

In order to carry out the study and demonstrate the results obtained, first, the authors consider a simple prototype for an anharmonic oscillator with cubic nonlinearity; that is,
\begin{equation}
\ddot{x}-\gamma x+\beta x^{3} = 0.
\label{eqn1}
\end{equation}
Here, the overdot denotes differentiation with respect to time, $\gamma$ is the coefficient of the linear stiffness of the oscillator, and $\beta$ is the coefficient (strength) of the cubic stiffness nonlinearity. Equation~(\ref{eqn1}) is said to preserve $\mathcal{P}$-symmetry; that is, $x = -x$. For the present numerical study, the authors have fixed the parameter values at $\gamma=0.5$ and $\beta=0.5$. The system (\ref{eqn1}) has three equilibrium points: $X_{0}=(0,0)$ and $X_{1,~2}=\left(\pm\sqrt{\gamma/\beta}, 0\right)$. Since the system has $\mathcal{P}$-symmetric property, for the chosen parameter values, the system has one saddle equilibrium point at ($0,0$) that is an unstable fixed point and two centers at ($\pm1,0$).  Henceforth, the local regions that include these points are referred to as positive and negative potential-wells, respectively. All three equilibrium points are depicted in Fig.~\ref{fig1}(a), in which the two centers are represented as open triangles and the saddle point is represented by a square. When one chooses initial conditions in either the positive well or the negative potential-well, the trajectories move in a clockwise direction and undergo single-well, small-amplitude periodic oscillations. On the contrary, for initial conditions that are chosen away from the equilibrium points, the system exhibits double-well, large-amplitude periodic oscillations. The orbit that separates the single-well and double-well oscillatory regions in the phase space is known as homoclinic orbit, which is plotted in Fig.~\ref{fig1}(a) with a black line. Different initial conditions are depicted as solid circles. The trajectories started from initial conditions within the homoclinic orbit result in single-well periodic oscillations and the trajectories started from outside the homoclinic orbit experience large excursions to both potential-wells resulting in double-well oscillations that are also illustrated in Fig.~\ref{fig1}(a). 
	
To understand the underlying dynamical mechanism, the authors have calculated the total energy of the system, which is given by,
\begin{equation}
E_1=\frac{1}{2}\left[\dot{x}^{2}+\frac{\beta}{2} x^{4}-{\gamma}x^{2}\right].
\label{eqn2}
\end{equation}
If one substitutes the initial values for $x$ and $\dot{x}$ into Eq.~(\ref{eqn2}), then $E_1$ has negative values for the initial conditions chosen within the homoclinic orbit and the trajectories started from these initial conditions remain inside and result in single-well periodic oscillations. Beyond this orbit, $E_1\geq0$, and the trajectories move away from the homoclinic orbit resulting in double-well oscillations as shown in Fig.~\ref{fig1}(a).
	
Next, the authors introduce a position dependent damping, called here as nonlinear damping of the form $\alpha x \dot{x}$ into the system (\ref{eqn1}). The resulting system is given by 
\begin{equation}
\ddot{x}+\alpha x\dot{x}-\gamma x+\beta x^{3} = 0,
\label{eqn3}
\end{equation}
which is of the Li\'enard type $\ddot{x}+f(x)\dot{x}+g(x)=0$, where $f(x)=\alpha x$ and $\alpha$ is the nonlinear damping coefficient, and $g(x)=-\gamma x+\beta x^{3}$ in Eq.~(\ref{eqn3}). The system (\ref{eqn3}) can be viewed as a cubic anharmonic oscillator (\ref{eqn1}) with nonstandard Hamiltonian nature \cite{chandru2005}, or as a conservative nonlinear oscillator perturbed by the  nonlinear damping $\alpha x \dot{x}$. For the past several years, the invariance and integrability properties of this equation have been studied in detail \cite{mahomed1985, mahomed1987, chandru2005a, duarte1987}. When one includes the nonlinear damping term, the $\mathcal{P}$-symmetry of the system (\ref{eqn1}) is broken instantly and Eq.~(\ref{eqn3}) has $\mathcal{PT}$--symmetric nature. That is, $x=-x$ and $t=-t$. 
	
As in the anharmonic oscillator (\ref{eqn1}), the system (\ref{eqn3}) has three equilibrium points at $X_{0}=(0,0)$ and $X_{1,~2}=\left(\pm\sqrt{\gamma/\beta}, 0\right)$. However, due to the $\mathcal{PT}$ -- symmetric property, for the chosen parameter values the two centers ($X_{1,~2}$) in the two potential-wells become stable focus ($X_{1}$) and unstable focus ($X_{2}$) in the corresponding positive and negative potential-wells, respectively. Positions of the saddle, stable and unstable focus equilibrium points are indicated in Fig.~\ref{fig1}(b) with a square, an open circle, and a triangle, respectively. 
	
The trajectories are attracted to the stable focus in the positive potential-well when the initial conditions are chosen within the domain of attraction of this equilibrium point.  For trajectories initiated outside this domain of attraction, the resulting motion is in the form of periodic oscillations. Based, on the choice of initial conditions, the system has either a dissipative or a non-dissipative nature. The black closed loop in Fig.~\ref{fig1}(b) is used to denote the homoclinic orbit that separates regions with different types of motions.  To be precise, one can consider the divergence of the vector field of system ~(\ref{eqn3}) expressed in a state-space form with the states being $x$ and $\dot{x}$.  This divergence is equal to $- \alpha x$, which is positive, negative, or zero depending on the location in state space. For a positive value of $\alpha$,  the system is dissipative when $x>0$, conservative when $x=0$, and neither conservative nor dissipative when $x<0$. The nonlinear damping term in Eq.~(\ref{eqn3}) acts as an energy dissipating term as well as an energy adding term, which can give rise to self-sustained oscillations. This position-dependent phenomenon enables one to understand the controlling aspects of certain biological and chemical oscillations \cite{ghosh2014}.
	
One can also examine the coexistence of dissipative and non-dissipative nature of the system (\ref{eqn3}) in terms of the total energy of the system \cite{suresh2018}. One can write down the total energy for the system given by Eq.~(\ref{eqn3}) as
\begin{eqnarray}
{E_{2}} = &\frac{1}{2}\left[\dot{x}^{2}+\frac{\alpha \dot{x}}{2}\left(x^{2}-\frac{\gamma}{\beta}\right)+\frac{\beta}{2}\left(x^{2}-\frac{\gamma}{\beta}\right)^{2}\right]\times ~e^{\frac{\alpha}{\Omega}\tan^{-1}\left[\frac{\alpha \dot{x}+2\beta\left(x^{2}-\frac{\gamma}{\beta}\right)}{2\Omega \dot{x}}\right]}-\left(\frac{\gamma^2}{4\beta}\right)e^{\frac{\alpha\pi}{2\Omega}},
\label{eqn4} 
\end{eqnarray}
where $\Omega=\frac{1}{2}\sqrt{8\beta-\alpha^{2}}$. If one substitutes the initial conditions for ($x,~\dot{x}$) into Eq.~(\ref{eqn4}), then, for some initial conditions, $E_2$ has negative values and for those locations in the phase space the system has a dissipative nature. On the other hand, the system has a non-dissipative nature, when the total energy of the system $E_{2}\geq0$. In particular, in Fig.~\ref{fig1}(b), the total energy of the system has negative values inside the homoclinic orbit. For phase space locations outside the homoclinic orbit, one has positive $E_{2}$ values. 
\begin{figure}
\centering \includegraphics[width=0.5\columnwidth]{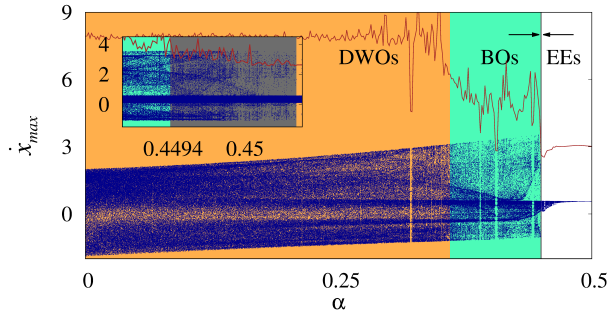}
\caption{One-parameter bifurcation diagram plotted by using the maxima of the system variable $\dot{x}$ from Eq.~(\ref{eqn5}) illustrating the occurrence of double-well oscillations (DWOs), BOs, EEs and single-well chaos as a function of $\alpha\in[0,0.5]$ with $F=0.2$ and $\omega=0.7315$. Other system parameters are fixed as the values used to generate Fig.~\ref{fig1}. The red (dark gray) line represents the threshold height $H_s$. In the inset, the presence of EEs is depicted.}
\label{fig2}
\end{figure}
	
It is recalled that with the addition of the nonlinear damping term $\alpha x\dot{x}$ in Eq.~(\ref{eqn1}), the $\mathcal{P}$ -- symmetry of the anharmonic oscillator is broken instantly and transformed into a $\mathcal{PT}$ -- symmetry. Consequently, the fixed point in the negative potential-well turns into a source and repels trajectories started within the homoclinic orbit. The repelled trajectories are attracted by the equilibrium point in the positive well that acts as a sink. When one quasi-statically increases the strength of nonlinear damping $\alpha$, apparently, the dissipation is quicker. However, it is really intriguing to understand the dynamics of the system when it exhibits chaotic behavior and how the symmetry breaking influences the chaotic dynamics of this system with respect to the variation in the nonlinear damping strength. To investigate this, in the next section, the authors include an external periodic force into the system (\ref{eqn3}) and study the development of new dynamical states that emerge with respect to the nonlinear damping. 
\begin{figure*}
\centering \includegraphics[width=1.0\columnwidth]{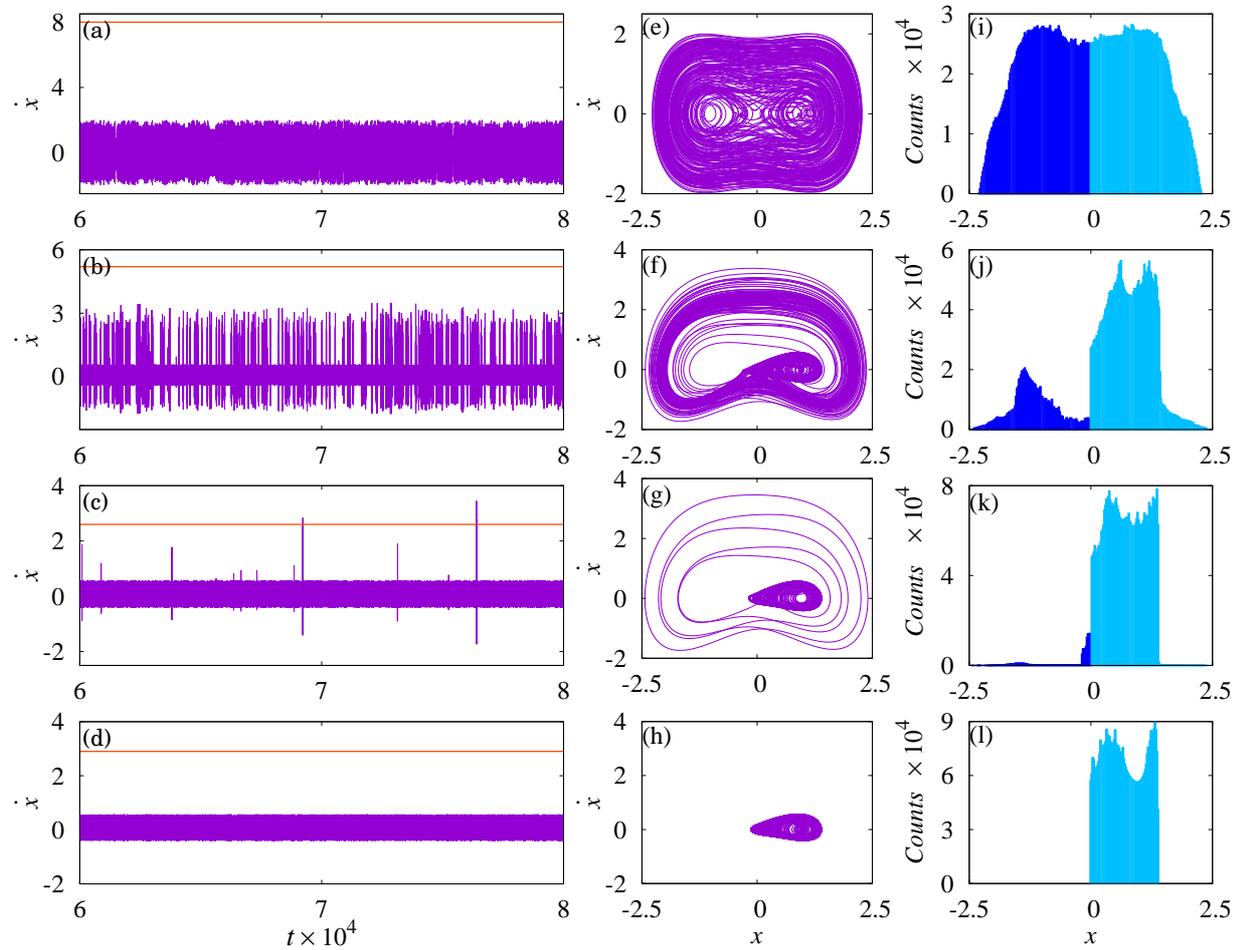}
\caption{Time evolution, phase portrait, and counts of the events that occurred in both potential-wells of the system (\ref{eqn5}). (a), (e) and (i): Double-well chaotic oscillations for $\alpha=0$. (b), (f) and (j): BOs for $\alpha=0.44$. (c), (g) and (k): EEs for $\alpha=0.45$. (d), (h) and (l):  Single-well bounded chaos for $\alpha=0.46$.}
\label{fig3}
\end{figure*}
\section{Dynamics of the forced anharmonic oscillator with nonlinear damping}
\label{sec3}
With the inclusion of an external periodic force of the form $F\sin(\omega t)$, Eq.~(\ref{eqn3}) can be rewritten as,
\begin{equation}
\ddot{x}+\alpha x\dot{x}-\gamma x+\beta x^{3} = F\sin(\omega t),
\label{eqn5}
\end{equation}
here $F$ and $\omega$ are the amplitude and frequency of the sinusoidal excitation, respectively. The integrable property and bifurcation structures of the system (\ref{eqn3}) have been previously studied in detail for $F=0$ \cite{chandru2007,karthiga2016}. When $F\neq0$ in Eq,~(\ref{eqn5}), the symmetry of the equilibrium points is broken and they start moving along the $x$ axis with respect to the forcing amplitude $F$ \cite{suresh2018}. Due to this oscillation of the equilibrium points, the dissipative and non-dissipative regions of the system are also oscillating in time. Consequently, the dissipative region is enlarged in the phase space. This oscillation of the fixed points is governed by stretching and folding actions, which eventually lead the system to chaotic behavior when the initial conditions are chosen from the dissipative region and for suitable values of $F$ and $\omega$. For the current study, the amplitude of the external force is fixed at $F=0.2$ units. In the absence of nonlinear damping (i.e., $\alpha=0$), with a suitable value of $\omega$, the system (\ref{eqn5}) exhibits double-well chaos. Here, the authors mention double-well chaos in the sense that the system jumps alternatively between the two potential-wells and averagely spends equal time in them. When the nonlinear damping is included, the system has an unstable focus in the negative potential-well and the trajectories are repelled by this equilibrium point. Consequently, time spent by the system in the negative potential-well gets reduced, and for most of the time, the system oscillates only in the right potential-well for sufficiently large values of $\alpha$. If one looks at this situation in the time domain, the system exhibit bounded small-amplitude (single-well) chaotic oscillations at most of the times and travels to the next potential-well intermittently, which produces large-amplitude (double-well) oscillations. This type of oscillation is generally known as bursting-like oscillations, during which the system exhibits coexisting large-amplitude oscillations alternating with the small-amplitude oscillations. The system exhibits BOs for a range of nonlinear damping parameter values. To classify BOs from double-well oscillations, the authors have calculated the total time (T) spent by the system in the right and left potential-wells, namely T$_{R}$ and T$_{L}$, respectively, and estimated the ratio T$_{ratio}$ = T$_{L}$/T$_{R}$.  For T$_{ratio}>$ 0.5, the oscillations are characterized as double-well oscillations; otherwise, they are quantified as BOs. If one were to increase $\alpha$ further, the large-amplitude chaotic bursts are found to occur occasionally (with the ratio of T$_{ratio}<$ 0.1) and randomly with highly unpredictable nature along with small-amplitude chaos. This dynamical state is separately characterized as EEs and the threshold height $H_s$ is used to distinguish it from the other dynamical states. Further, at a critical value of $\alpha$, the large-amplitude oscillations are suddenly reduced and the system exhibit single-well chaos with the ratio of T$_{ratio}=0$, which then eventually leads to periodic oscillations (via reverse period-doubling bifurcation) for larger values of $\alpha$. One can note that when $\alpha$ has negative values, then the trajectories are attracted into the left potential-well. 
	
To verify this transition, the authors have numerically calculated the one-parameter bifurcation diagram of the system by plotting the maxima of the dynamical variable $\dot{x}$ of the system (\ref{eqn5}) as a function of $\alpha$, as shown in Fig.~\ref{fig2}. In the range of $\alpha\in[0,0.365]$ (marked as orange (gray) region in Fig.~\ref{fig2}), the system oscillates alternatively in the two potential-wells resulting in double-well chaotic oscillations. The corresponding time evolution and the phase portrait are plotted in Figs.~\ref{fig3}(a) and \ref{fig3}(e), respectively, for $\alpha=0$ validating the double-well oscillations. Additionally, the count of the maxima in each bin (well) of the system variable $x$ is also calculated and plotted in Figs.~\ref{fig3}(i). One can discern that the number of events (oscillations) occurring in the two potential-wells are equal with the ratio of T$_{ratio}$ = 0.989, corroborating the double-well chaotic oscillations. When one increases the nonlinear damping strength further, for the range of $\alpha\in(0.365, 0.4493]$, the system exhibits BOs wherein the large-amplitude oscillations alternated with small-amplitude chaos. This region of BOs is marked in light green (gray) in Fig.~\ref{fig2}. The corresponding time series and phase portraits are plotted in Fig.~\ref{fig3}(b) and Fig.~\ref{fig3}(f), respectively, for $\alpha=0.44$. Compared to the double-well oscillations, in BOs, the number of maxima in the left potential-well is slightly reduced with T$_{ratio}$ = 0.435. This is evident from Fig.~\ref{fig3}(j). Furthermore, for $\alpha\in(0.4493, 0.4506]$ the system is found to display EEs where the large-amplitude oscillations occur occasionally and randomly along with the small-amplitude oscillations. The range of $\alpha$ for which the EEs occur is highlighted in dark gray and plotted separately as an inset in Fig.~\ref{fig2}. The time evolution and phase portrait plots are depicted in Figs.~\ref{fig3}(c) and \ref{fig3}(g), respectively, for $\alpha=0.45$.  These plots confirm the occasional occurrence of large-amplitude oscillations.  The authors also emphasize here that some of the large-amplitude oscillations are higher than the threshold $H_s$ (horizontal line), which are qualified as EEs, whereas in Figs.~\ref{fig3}(a) and \ref{fig3}(b) the threshold $H_s$ is larger than the system amplitude. Furthermore, in Fig.~\ref{fig3}(k), the authors show that the counts of maxima in the left potential-well are drastically reduced, while that the number of events in the right potential-well is increased with the ratio of T$_{ratio}$ = 0.071, which is also validating of the occurrence of EEs. For $\alpha>0.4506$, the system exhibits only single-well chaos.  The occurrence of single-well chaos is again confirmed from Fig.s.~\ref{fig3}(d), \ref{fig3}(h) and \ref{fig3}(l), which are plotted for $\alpha=0.46$. For larger values of $\alpha$, eventually, the system gives rise to periodic oscillations via reverse period-doubling bifurcation. The authors wish to point out here that the above mentioned dynamical states arise only if one chooses the initial conditions inside the dissipative region. If one chooses the initial conditions outside the dissipative region, the system exhibits quasi-periodic oscillations \cite{kingston2017a, suresh2018}.
	
Therefore, when one incorporates the nonlinear damping in the forced anharmonic oscillator and increases the damping strength, the $\mathcal{P}$--symmetry of the system is broken instantly, and the system has $\mathcal{PT}$ -- symmetry, which alters the stability of the equilibrium points. Consequently, the trajectories approaching the left potential-well are repelled by the unstable fixed point and attracted to the stable equilibrium point in the right potential-well. In other words, the total time spent by the system in the left (right) potential-well is decreased (increased) when one increases the damping strength, which manifests a transition from double-well to bursting-like oscillations and then to single-well chaos via EEs. Through this study, the authors have elucidated the origin and emerging mechanism of EEs in a forced anharmonic oscillator in the presence of nonlinear damping and external forcing. With regard to control of EEs, the studies carried out in the literature have been quite limited and narrow in scope. Therefore, this needs further attention. In this vein, in the next section, the authors introduce linear damping in Eq.~(\ref{eqn5}) and study the impact of it on EEs in the system response.
\begin{figure*}
\centering
\centering \includegraphics[width=1.0\columnwidth]{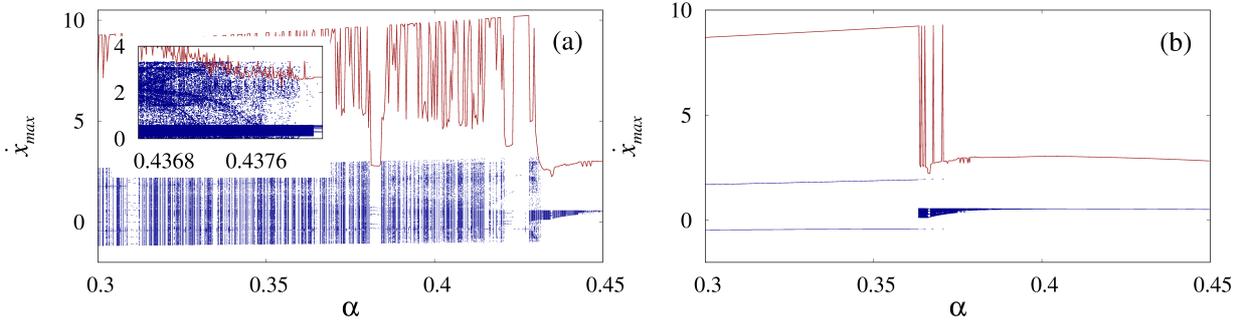}
\caption{Bifurcation diagram for responses of system ~(\ref{eqn6}) with respect to variation in the nonlinear damping strength ($\alpha$) for two different values of $\xi$ with $F=0.2$ and $\omega=0.7315$. (a) Occurrence of BOs and EEs for $\xi = 0.01$. (b) Existence of periodic oscillations and single-well chaos without the occurrence of EEs for $\xi = 0.06$. Red (dark gray) continuous line represents the threshold $H_s$.}
\label{fig4}
\end{figure*}
\section{Influence of linear damping on extreme events in forced anharmonic oscillator with nonlinear damping}
\label{sec4}
After including a linear damping term ($\xi\dot{x}$) in Eq.~(\ref{eqn5}), the resulting system is of the form
\begin{equation}
\ddot{x}+\alpha x\dot{x}+\xi \dot{x}-\gamma x+\beta x^{3} = F\sin(\omega t),
\label{eqn6}
\end{equation}
where $\xi$ is the strength of the linear damping, which is positively valued. If one considers the divergence of the vector field of system ~(\ref{eqn3}) expressed in state-space form with the states being $x$, $\dot{x}$, and $\theta = \omega t$, then the divergence is equal to $-(\alpha x + \xi)$.  It is clear from this expression that through a choice of an appropriately large enough value of $\xi$, the divergence can always be negative. This means that the flow can be dissipative throughout the phase space.  In keeping with this, if one increases the value of $\xi$ from zero, it is found that the large-amplitude oscillations are completely eliminated from the system dynamics even for small values of $\xi$ and only small-amplitude single-well chaos is feasible in the system (\ref{eqn6}). To confirm this phenomenon, the authors have plotted the one-parameter bifurcation diagram for the response of the system ~(\ref{eqn6}) with respect to $\alpha$ for two different fixed values of $\xi$. Figs.~\ref{fig4}(a) and \ref{fig4}(b) are for $\xi=0.01$ and $\xi=0.06$, respectively. The continuous line in Figs.~\ref{fig4}(a) and  \ref{fig4}(b) represents the threshold $H_s$. In the absence of linear damping, the system exhibit different dynamical states as shown in Fig.~\ref{fig2}. With the inclusion of linear damping with $\xi=0.01$, for low values of $\alpha$, the system exhibits double-well periodic oscillations and the system exhibits BOs in the range of $\alpha\in[0.3, 0.4369]$ with intermediate windows of periodic oscillations. Due to these periodic windows, the fluctuation in the threshold $H_s$ appears, as evident from Fig.~\ref{fig4}(a). Upon increasing the nonlinear damping strength in the range of $\alpha\in(0.4369, 0.4381]$, the system exhibits EEs, which is clearly discernible in the inset of Fig.~\ref{fig4}(a) wherein the amplitude of the system response is larger than the threshold $H_s$. For further increase in $\alpha$,  the system response is in the form of single-well chaotic oscillations. When the authors increase the linear damping strength to $\xi=0.06$, the large-amplitude oscillations are completely eradicated and the system exhibits periodic and single-well chaos with respect to variation in $\alpha$. This situation is illustrated in Fig.~\ref{fig4}(b) where the transition from double-well periodic state to single-well chaos occurs without EEs. 
	
\begin{figure*}
\centering
\centering \includegraphics[width= 1.0\columnwidth]{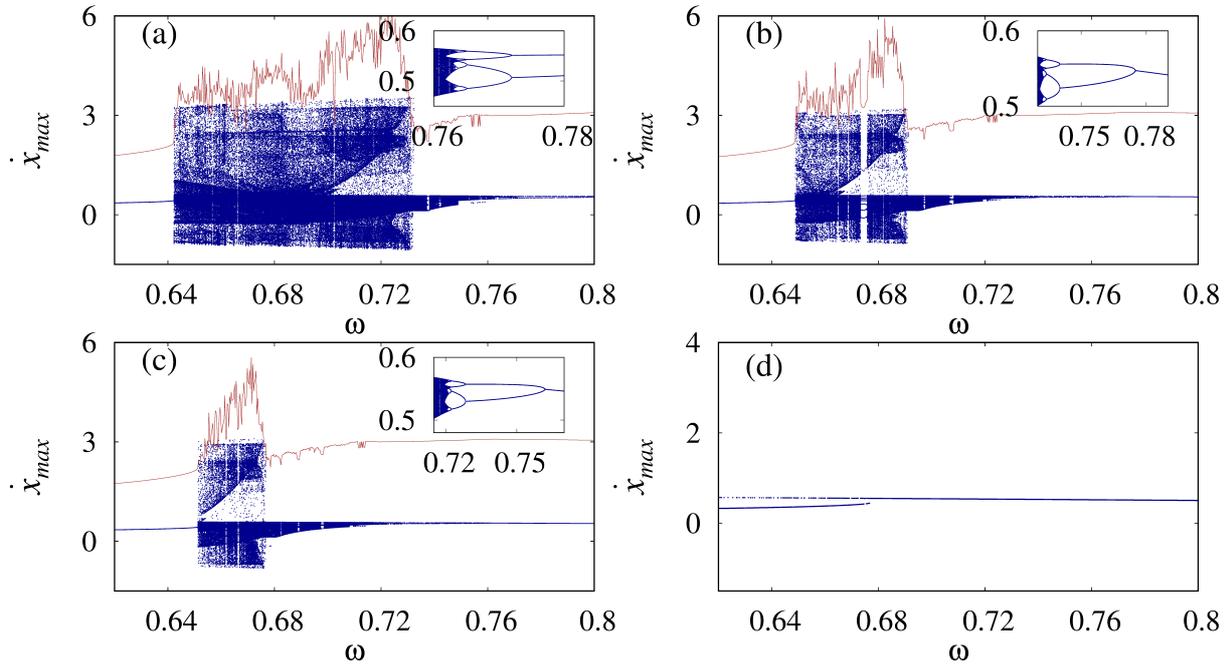}
\caption{(a-f) Bifurcation diagram for responses of ~(\ref{eqn6}) with respect to variation in the forcing frequency ($\omega$) for different values of $\xi$, $\alpha=0.45$, and $F=0.2$. (a) $\xi = 0$; there are two different routes for the emergence of EEs. (b) $\xi$ = 0.015, (c) $\xi$ = 0.02 and (d)  $\xi$ = 0.06; periodic oscillations occur in the considered region of $\omega$. Insets are included to depict the period-doubling sequences, which lead to single-well chaos when decreasing $\omega$.}
\label{fig5}
\end{figure*}

Accordingly, the large-amplitude oscillations are completely eliminated when one includes the linear damping into the forced anharmonic oscillator with nonlinear damping. It is also of interest to study how the large oscillations are removed from the system for a fixed value of $\alpha$.  To this end, the value of $\alpha$ is fixed as constant and the authors study the system dynamics by varying the forcing frequency $\omega$ and linear damping strength $\xi$.  In the absence of linear damping ($\xi=0$) in Eq.~\ref{eqn5}, large chaotic bursts occur via two distinct dynamical routes, namely, the intermittency and period-doubling routes as one see the Fig.~\ref{fig5}(a) from left to right by increasing the forcing frequency ($\omega$) from lower values and observe the figure from right to left by decreasing $\omega$ from higher values, respectively. This dynamics is depicted in Fig.~\ref{fig5}(a), in which the maxima of the system variable ($\dot{x}$) is plotted with respect to the forcing frequency ($\omega$) for $\xi=0$. In this figure, if one moves from left to right by increasing  $\omega$, the system undergoes a sudden transition from periodicity to large-amplitude chaotic bursting via the intermittency route at $\omega=0.64225$. On the other hand, if one examines the results provided in Fig.~\ref{fig5}(a), as one goes from right to left by decreasing $\omega$, the periodic attractor bifurcates into bounded single-well chaos via the period-doubling bifurcation sequence, which can be seen from the inset of Fig.~\ref{fig5}(a). When decreasing $\omega$ further, the chaotic attractor slowly increases in size and suddenly at $\omega=0.7316$ it expands into a large size attractor associated with EEs. In Fig.~\ref{fig5}(a), in the range of $\omega\in[0.64225, 0.7316]$, the system exhibits large-amplitude oscillations. The regions where EEs occur are marked with black points and the ones where BOs occur are marked with gray points. 
	
For non-zero values of $\xi$ in Eq.~(\ref{eqn6}), the occurrence of large-amplitude oscillations in the $\omega$ parameter region is significantly reduced. This case is illustrated in Figs.~\ref{fig5}(b) to \ref{fig5}(d) for different values of $\xi$. In the absence of $\xi$, large-amplitude oscillations occur in the range of $\omega\in[0.64225, 0.7316]$, whereas for $\xi=0.015$, this range is drastically reduced to $\omega\in[0.6488, 0.6906]$. Upon increasing the strength of linear damping to $\xi=0.02$, the region of large-amplitude oscillations is further reduced and occurs only in a small portion of $\omega\in[0.6513, 0.6767]$ as depicted in Fig.~\ref{fig5}(c). Finally, for $\xi=0.06$, the system exhibits only periodic oscillations, which is shown in Fig.~\ref{fig5}(d). From these results, one can confirm that the large-amplitude oscillations are completely removed from the system when one includes and increases the linear damping strength. 
	
\begin{figure}
\centering
\centering \includegraphics[width= 0.5\columnwidth]{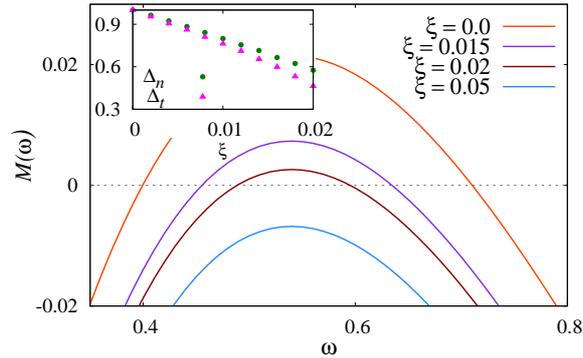}
\caption{ Melnikov function $M{(\omega)}$ variation with respect to the excitation frequency $\omega$ for different $\xi$ values.  One can note that the window over this function is positive is reduced as $\xi$ is increased.}
\label{fig6}
\end{figure}

The reduction of the excitation frequency window over which chaos is possible in the system (\ref{eqn6}) is analytically confirmed by computing the Melnikov function \cite{han2012}. This function is an analytical tool that can be used to study the global behavior of the system. Specifically, this function provides a procedure for analyzing and estimating when chaotic behavior is expected in the system. In order to be able to carry out the Melnikov analysis, one needs to consider the external forcing ($F$) and the nonlinear damping ($\alpha$) terms in Eq.~(\ref{eqn6}) as small perturbations. The Melnikov function associated with Eq.~(\ref{eqn6}) is given by
\begin{equation}
M(\omega)=\frac{\sqrt{\beta}\left(12\sqrt{2}\pi\beta F\Omega\Lambda-\gamma\left(3\pi\sqrt{2}\alpha\gamma+16\beta\xi  \sqrt{\frac{\gamma}{\beta}}\right)\right)}{12\beta^2},
\label{eqn7}
\end{equation}
%
where $\Lambda=\text{sech}\left(\frac{\pi  \omega}{2\sqrt{\gamma}}\right)$. When  $M(\omega)$ is positive, then the system can exhibit chaotic dynamics for the corresponding parameter values.  In Fig.~\ref{fig6}, the authors have plotted the Melnikov function $M(\omega)$ with respect to the forcing frequency $\omega$ for different values of $\xi$ (same values as those used to generate the results of Fig.~\ref{fig5}). It is noted that $M(\omega)$ is positive valued in a certain range of $\omega$ for $\xi=$ 0.0, 0.015, and 0.02, whereas for $\xi$ = 0.05, $M(\omega)<0$ for the entire region of $\omega$, which is in conformity with the authors' numerical findings. 	
	
It is noted here that the analytically calculated chaotic region in terms of the  $\omega$ parameter does not exactly match with the numerically obtained results given in Fig.~\ref{fig5}, since it has been assumed for the Melnikov function calculation that the external forcing and the nonlinear damping terms in Eq.~(\ref{eqn6}) are small perturbations. However, the widths of the chaotic regions obtained both numerical and theoretical calculations are in good agreement. To validate this, the authors have calculated the difference $\Delta_{0}=(\omega_{2}^{0}-\omega_{1}^{0})$ for $\xi=0$ and $\Delta_{\xi}=(\omega_{2}^{\xi}-\omega_{1}^{\xi})$ for different values of $\xi$, where $\omega^{0,\xi}_{2}$ and $\omega^{0,\xi}_{1}$ are the critical values of the forcing frequency at which chaos emerged via the period-doubling bifurcation and intermittency routes, respectively. The ratio $\Delta = \Delta_{\xi}/\Delta_{0}$ is estimated for both numerically obtained data ($\Delta_{n}$) from the maximal Lyapunov exponent of Eq.~(\ref{eqn6}) and theoretically obtained results ($\Delta_{t}$) from the Melnikov function (\ref{eqn7}).  The comparison between these results is shown in the inset of   Fig.~\ref{fig6}.  There is good agreement with each other.
	
\begin{figure}
\centering \includegraphics[width= 1.0\columnwidth]{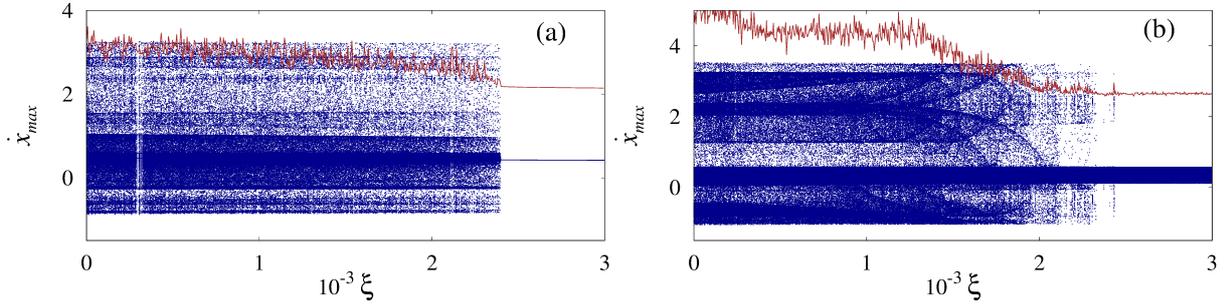}
\caption{Bifurcation diagram for the system (\ref{eqn6}) obtained with the control parameter being the linear damping ($\xi$). (a) $\omega=0.6432$; system undergoes a transition from EEs to periodic oscillations via intermittency route. (b)  $\omega=0.7263$; transition from BOs to single-well chaos mediated by EEs.}
\label{fig7}
\end{figure}

From Fig.~\ref{fig5}, the authors found that the EEs emerged via two different routes with respect to variation in the forcing frequency.  It is also of interest to understand how the elimination of such large events occurred as a function of linear damping strength ($\xi$) in these two different routes. To this end, the authors have again plotted the bifurcation diagrams for the responses of Eq.~(\ref{eqn6}) with respect to variation in $\xi$ for $\omega=0.6432$ (near intermittency route) and for $\omega=0.7263$ (near period-doubling route), as depicted in Figs.~\ref{fig7}(a) and \ref{fig7}(b), respectively. In Fig.~\ref{fig7}(a), in the absence of linear damping ($\xi=0$), the system exhibits EEs in which the amplitude of the oscillations is higher than the threshold $H_s$. After including the linear damping and increasing its strength for sufficiently large values, the system undergoes a transition from EEs to periodic oscillations. On the other hand, for $\omega=0.7263$, in the absence and for low values of linear damping, the system exhibits BOs. When one increases the damping strength, for large values, EEs still occur in the system. As one further increases $\xi$, the system response is led to single-well chaotic oscillations. Hence, the authors believe that they have identified that the elimination of EEs can occur through two distinct routes depending on the values of $\omega$ and $\xi$.
	
\begin{figure}
\centering \includegraphics[width=0.5\columnwidth]{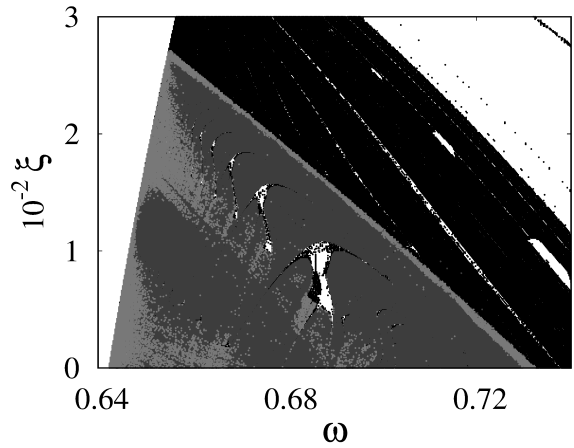}
\caption{Elimination of large-amplitude oscillations, such as BOs and EEs from the system (\ref{eqn6}) in the control parameter space spanned by the external forcing frequency ($\omega$) and linear damping strength ($\xi$).  In the light gray colored region, EEs occur.  BOs occur in the dark gray colored region. The black-colored region corresponds to single-well bounded chaos and the region in white is where periodic oscillations occur.}
\label{fig8}
\end{figure}

To identify the global dynamical behavior of the system and to understand the elimination of large events in a large parameter plane, the authors have numerically studied the two-parameter space by varying the forcing frequency ($\omega$) and linear damping strength ($\xi$), as illustrated in Fig.~\ref{fig8}. The regions corresponding to EEs, BOs, single-well chaotic motions, and periodic motions are shown in this figure. To separate the BOs and EEs, the authors use the threshold $H_s$. When the peak response values are larger than $H_s$, the corresponding motion is labelled as an EE. On the other hand, when the peak response values are smaller than $H_s$, the corresponding motion is labelled as a MMO. The largest Lyapunov exponent of the system is calculated to distinguish chaos from periodic oscillations. From Fig.~\ref{fig8}, one can clearly note that there are two distinct routes to the elimination of large-amplitude chaotic oscillations with respect to the parameter $\xi$. It is also worth noting here that when one increases the strength of linear damping, the transition from EEs to periodic oscillations occurs over a narrow region of $\omega$, whereas the transition from EEs to single-well chaos occurs over a wide range of the $\omega$ plane which is also evident from Fig.~\ref{fig8}. 
	
Based on the results presented here, it is believed that the authors have identified that the large-amplitude oscillations can be completely eliminated from the nonlinearly damped and forced anharmonic oscillator with an appropriate strength of linear damping. In the next section, the possible mechanism for the elimination of large-amplitude oscillations is examined.
%
\begin{figure}
\centering \includegraphics[width=0.5\columnwidth]{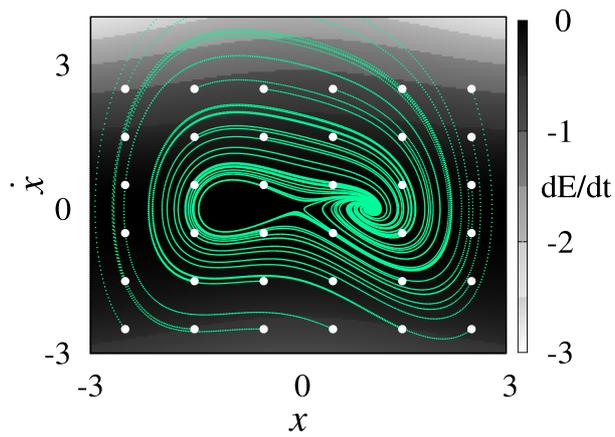}
\caption{Phase portrait of Eq.~(\ref{eqn6}) with $F=0$. Evolutions of trajectories that emerge from different initial conditions are illustrated in green (light gray) lines. The chosen initial conditions are marked with white filled circles. The rate of change of energy $\frac{dE}{dt}$ is depicted with a gray shade.  One can note that the rate is negative valued, confirming that the system (\ref{eqn6}) has a decaying solution or a dissipative behavior in the whole phase space.}
\label{fig9}
\end{figure}
\section{A mechanism for controlling extreme events}
\label{sec4a}
The inclusion of linear damping into Eq.~(\ref{eqn6}) is found to transform the non-dissipative nature of the system into a dissipative one.  This occurs throughout the phase space. To illustrate this, the authors have calculated the change in the total energy of the system (\ref{eqn6}) in the presence of linear damping and nonlinear damping without external forcing ($F=0$). The rate of change of energy $\left(\frac{dE}{dt}\right)$ of Eq.~\ref{eqn6} can be written as
\begin{eqnarray}
\frac{dE}{dt} = &-e^{\frac{\alpha}{\Omega}\tan^{-1}\left[\frac{\alpha \dot{x}+2\beta\left(x^{2}+\frac{\gamma}{\beta}\right)}{2\Omega \dot{x}}\right]}\times (\xi\dot{x}^{2}) < 0,
\label{eq8} 
\end{eqnarray}
where $\frac{dE}{dt}<0$ indicates that the system has a decaying solution or a dissipative nature.  Due to this dissipative nature, the trajectories started from anywhere in the phase space have an exponentially decaying solution as a function of time and converge towards the stable focus in the right potential-well. To elucidate this, the authors have plotted the phase space diagram with trajectories started from different initial conditions (inside and outside of the homoclinic orbit) for $\xi=0.1$ in Fig.~\ref{fig9}. It is evident from this figure that all of the trajectories started from different initial conditions slowly converge towards the stable focus. Nevertheless, without the linear damping, the trajectories started outside the homoclinic orbit have neutrally stable periodic orbits, as depicted in Fig.~\ref{fig1}(b). The authors have also plotted the rate of change of energy $\left(\frac{dE}{dt}\right)$ in Fig.~\ref{fig9} as a gray shaded.  It is pointed out that if one includes the external periodic forcing with suitable values of $F$ and $\omega$, then the system exhibits chaotic oscillations which are confined within the single-well without having any large excursions (double-well oscillations). As mentioned earlier, the divergence of the vector field of the forced system can be kept negative through an appropriate choice of the linear damping strength $\xi$.  From these observations, one can realize that the inclusion of linear damping can modify the system to be dissipative throughout the whole phase space.  This is found to be a key to the elimination of large-amplitude oscillations such as BOs and EEs. 
\section{Conclusions}
\label{sec5}
To close the article, it is stated that the authors have studied the dynamics of a forced anharmonic oscillator by including nonlinear damping and linear damping terms. The unforced anharmonic oscillator has $\mathcal{P}$-symmetry. This means that the equilibrium points in both positive and negative potential-wells are identical. The system exhibit single-well periodic oscillations if one chooses the initial conditions very near to the neutrally stable elliptic points (in both the wells), whereas the system exhibit double-well periodic oscillations when one chooses the initial conditions away from the equilibrium points. These different motions can be distinguished by using the total energy of the system, as discussed here.
	
Furthermore, the author have studied the effect of nonlinear damping in the anharmonic oscillator by including the nonlinear damping term $x\dot{x}$ and found that the system becomes $\mathcal{PT}$-symmetric nature due to the presence of nonlinear damping term. Hence, the stable nature of the equilibrium points is changed and the two neutrally stable equilibrium points become unstable and stable focus in the negative and positive potential-wells, respectively. This system is shown to be capable of generating two distinct dynamical behaviors (dissipative and non-dissipative) depending on the initial conditions. The total energy of the system has been derived to illustrate the dual nature of the system.  Additionally, the authors have shown that under the influence of an external periodic force, without the nonlinear damping term, the system exhibits double-well chaotic behavior for certain values of amplitude and frequency of the forcing. If one includes the nonlinear damping and increases the damping strength, for large values, the system undergoes a transition from double-well chaotic oscillations to single-well chaos mediated by EEs. 
	
To influence the dissipation characteristics of the system, the linear damping is included into the forced anharmonic oscillator along with nonlinear damping. This inclusion if found to help in completely eliminating large-amplitude events from the system dynamics.  In the control parameter space spanned by the forcing frequency and the strength of linear damping, the authors have identified that the elimination of such large-amplitude oscillations occur through two distinct routes, one, a transition from EEs to periodic oscillations, and another, a transition from BOs to single-well chaos intervened by EEs. These results have been supported both numerically by plotting the bifurcation diagrams and analytically by calculating the Melnikov function. The analytically determined results are found to agree well with the numerically obtained results. The mechanism for the elimination of EEs has also been examined with numerical studies and the findings are in good agreement with the authors' analytically obtained results. By including the linear damping, the authors illustrate that one can realize a system with a dissipative nature throughout the phase space.  This is found to be a key for the suppression of EEs.  It is emphasized that the results shown in this work are robust to changes in the system and forcing parameters. 
	
\section{Acknowledgment}
B. Kaviya acknowledges SASTRA Deemed University for providing Teaching Assistantship. The work of R. Suresh is supported by the SERB-DST Fast Track scheme for Young Scientist under Grant No. YSS/ 2015/001645. The work of V. K. Chandrasekar forms a part of a research project sponsored by the CSIR EMR Grant No. 03(1444)/18/ EMR-II and B. Balachandran gratefully acknowledges the partial support received for this work through the U.S. National Science Foundation Grant No. CMMI1854532
	
	
	
	

\begin{thebibliography}{99}
		%
		\bibitem{guckenheimer1983}
		Guckenheimer J, Holmes PJ. Nonlinear oscillations, dynamical systems and bifurcation of vector fields. New York: Springer; 1983.
		%
		\bibitem{kovacic2011}
		Kovacic I,  Brennan MJ. The Duffing Equation: Nonlinear Oscillators and their Behaviour. London: Wiley; 2011.
		%
		\bibitem{lakshmanan1996}
		Lakshmanan M, Murali K. Chaos in Nonlinear Oscillators: Synchronization and Control. World Scientific. Singapore: 1996.
		%
		\bibitem{ji2002}
		Ji JC, Leung AYT. Bifurcation control of a parametrically excited Duffing system. Nonlinear Dyn 2002;27:411-17. 
		%
		\bibitem{li2006}
		Li H, Preidikman S, Balachandran B, Mote Jr  CD. Nonlinear free and forced oscillations of piezoelectric microresonators. J. Micromechanics and Microengineering 2006;16:356-67.
		%
		\bibitem{eichler2011}
		Eichler A, Moser J, Chaste J, Zrdojek M, Wilson-Rae I, Bachtold  A. Nonlinear damping in mechanical resonators made from carbon nanotubes and graphene. Nat. Nanotechnol 2011;6(339). 
		%
		\bibitem{ekinci2004}
		Ekinci KL, Yang YT, Roukes ML. Ultimate limits of inertial mass sensing based upon nanoelectromechanical systems. J. Appl. Phys 2004;95:2682-9.
		%
		\bibitem{papariello2016}
		Papariello L, Zilberberg O, Eichler A, Chitra A. Ultrasensitive hysteretic force sensing with parametric nonlinear oscillators. Phys. Rev. E 2016;94(022201). 
		%
		\bibitem{akerman2010}
		Akerman N, Kotler S, Glickman Y, Dallal Y, Keselman A, Ozeri R. Single-ion nonlinear mechanical oscillator. Phys. Rev. E 2010;82(061402(R)).
		%
		\bibitem{zaitsev2012}
		Zaitsev S, Shtempluck O, Buks E, Gottlieb O. Nonlinear damping in a micromechanical oscillator. Nonlinear Dyn 2012;67:859-83.
		%
		\bibitem{ran2009}
		Ran Z. One exactly soluble model in isotropic turbulence. Appl. Fluid Mech 2009;5:41-67.
		%
		\bibitem{siewe2009}
		Siewe MS, Cao H, Sanju\'an  MAF. Effect of nonlinear dissipation on the basin boundaries of a driven two-well Rayleigh–Duffing oscillator. Chaos Solitons Fractals 2009;39:1092-99.
		%
		\bibitem{miwadinou2015}
		Miwadinou CH, Monwanou AV, Chabi Orou JB. Effect of Nonlinear Dissipation on the Basin Boundaries of a Driven Two-Well Modified Rayleigh–Duffing Oscillator. Int. J. Bifurc. Chaos 2015;25(1550024).
		%
		\bibitem{ravindra1995}
		Ravindra B, Mallik AK. Chaotic response of a harmonically excited mass on an isolator with non-linear stiffness and damping characteristics. J. Sound Vib 1995;182:345-53. 
		%
		\bibitem{ravindra1994}
		Ravindra B, Mallik AK. Role of nonlinear dissipation in soft Duffing oscillators. Phys. Rev. E 1994;49:4950-54. 
		%
		\bibitem{ravindra1994a}
		Ravindra B, Mallik AK. Stability analysis of a nonlinearly damped Duffing oscillator. J. Sound Vib 1994;171:708-16. 
		%
		\bibitem{bikdash1994}
		Bikdash MU, Balachandran B, Nayfeh AH. Melnikov analysis for a ship with general roll damping, Nonlinear Dyn 1994;6:101-24. 
		%
		\bibitem{nayfeh1995}
		Nayfeh AH, Mook DT. Nonlinear Oscillations. New York: Wiley (Wiley Classics Library); 1995. 
		%
		\bibitem{almog2007}
		Almog R, Zaitsev S, Shtempluck O, Buks E. Noise Squeezing in a Nanomechanical Duffing Resonator. Phys. Rev. Lett 2007;98(078103). 
		%
		\bibitem{baltanas2001}
		Baltanas JP, Trueba JL, Sanju\'an MAF. Energy dissipation in a nonlinearly damped Duffing oscillator. Physica D 2001;159:22-34.
		%
		\bibitem{sanjuan1999}
		Sanju\'an MAF. The effect of nonlinear damping on the universal escape oscillator. Int. J. Bifurc. Chaos 1999;9:735-44.
		%
		\bibitem{jing2009}
		Jing XJ, Lang, ZQ. Frequency domain analysis of a dimensionless cubic nonlinear damping system subject to harmonic input. Nonlinear Dyn 2009;58:469-85.
		%
		\bibitem{leuch2016}
		Leuch A, Papariello L, Zilberberg O, Degen CL, Chitra R, Eichler A. Parametric Symmetry Breaking in a Nonlinear Resonator. Phys. Rev. Lett 2016;117(214101).
		%
		\bibitem{lifshitz2008}
		Lifshitz R, Cross MC. Nonlinear Dynamics of Nanomechanical and Micromechanical Resonators. New York: Wiley; 2008.
		%
		\bibitem{patidar2016}
		Patidar V, Sharma A, Purohit G. Dynamical behaviour of parametrically driven Duffing and externally driven Helmholtz–Duffing oscillators under nonlinear dissipation. Nonlinear Dyn 2016;83:375-88.
		%
		\bibitem{kingston2017}
		Kingston KL, Tamilmaran K. Bursting oscillations and mixed-mode oscillations in driven Li\'enard system. Int. J. Bifurc. Chaos 2017;27(1730025).
		%
		\bibitem{kingston2017a}
		Kingston KL, Tamilmaran K, Pal P, Feudel U, Dana SK. Extreme events in the forced Li\'enard system. Phys. Rev. E 2017;96(052204).
		%
		\bibitem{kingston2018}
		Kingston KL, Suresh K, Tamilmaran K. Mixed-mode oscillations in memristor emulator based Li\'enard system. AIP Conference Proceedings 2018;1942(060008).
		%
		\bibitem{chandru2007}
		Chandrasekar VK, Senthilvelan M, Lakshmanan M. On the general solution for the modified Emden-type equation $\dddot{x}+\alpha x\dot{x}+\beta x^{3}=0$. J. Phys. A: Math. Theor. 2007;40(4717).
		%
		\bibitem{karthiga2016}	Karthiga S, Chandrasekar VK, Senthilvelan M, Lakshmanan M. Twofold PT symmetry in nonlinearly damped dynamical systems and tailoring PT regions with position-dependent loss-gain profiles. Phys. Rev. A 2016;93(012102).
		%
		\bibitem{suresh2018}
		Suresh R, Chandrasekar VK. Influence of time-delay feedback on extreme events in a forced Li\'enard system. Phys. Rev. E 2018;98(052211).
		%
		\bibitem{suresh2019}
		Suresh R, Chandrasekar VK. Parametric excitation induced extreme events in nonlinear systems. (Unpublished)
		%
		\bibitem{dysthe2008}
		Dysthe K, Krogstad HE, M\"uller P. Oceanic rogue waves. Annu. Rev. Fluid Mech 2008;40: 287-310.
		%
		\bibitem{chabalko2014}
		Chabalko C, Moitra A, Balachandran B. Rogue waves: new forms enabled by GPU Computing. Physics Letters A 2014;378:2377-81.
		%
		\bibitem{solli2007}
		Solli DR, Ropers C, Koonath P, Jalali B. Optical rogue waves. Nature 2007;450:1054-8.
		%
		\bibitem{chen2015}
		Chen Y-Z, Huang Z-G, Zhang H-F, Eisenberg D, Seager TP, Lai Y-C. Extreme events in multilayer, interdependent complex networks and control. Sci. Rep 2015;5(17277).
		%
		\bibitem{lehnertz2008}
		Lehnertz K. Epilepsy and Nonlinear Dynamics. J. Biol. Phys 2008;34:253-66.
		%
		\bibitem{bialonski2016}
		Bialonski S, Caron DA, Schloen J, Feudel U, Kantz H, Moorthi SD. Phytoplankton dynamics in the Southern California Bight indicate a complex mixture of transport and biology. J. Plankton Res 2016;38:1077-91.
		%
		\bibitem{dobson2007}
		Dobson I, Carreras BA, Lynch VE, Newman D.E. Complex systems analysis of series of blackouts: Cascading failure, critical points, and self-organization. Chaos 2007;17(026103).
		%
		\bibitem{chowdhury2019}
		Chowdhury SN, Majhi S, Ozer M, Ghosh D, Perc M. Synchronization to extreme events in moving agents. New J. Phys. 2019(073048).
		%
		\bibitem{ray2019}
		Ray A, Rakshit S, Ghosh D, Dana SK.Intermittent large deviation of chaotic  trajectory in Ikeda map: Signature of extreme events. Chaos 2019;29(043131).
		%
		\bibitem{chen2014}
		Chen Y-Z, Huang Z-G, Lai Y-C. Controlling extreme events on complex networks. Sci. Rep. 2014;4(6121).
		%
		\bibitem{cavalcante2013}
		Cavalcante HLDde S, Ori\'a M, Sornette D, Ott E. Predictability and Suppression of Extreme Events in a Chaotic System. Phys. Rev. Lett 2013;111(198701).
		%
		\bibitem{galuzio2014}
		Galuzio PP, Viana RL, Lopes SR. Control of extreme events in the bubbling onset of wave turbulence. Phys. Rev. E 2014;89(040901(R)).
		%
		\bibitem{han2012}
		Han M, Yu P. Fundamental theory of the Melnikov function method. In: Melnikov Functions and Bifurcations of Limit Cycles page numbers. London: Springer; 2012.
		%
		\bibitem{chandru2005}
		Chandrasekar VK, Senthilvelan M, Lakshmanan M. Unusual Li\'enard-type nonlinear oscillator. Phys. Rev. E 2005;72(066203). 
		%
		\bibitem{mahomed1985}
		Mahomed FM, Leach PGL. The linear symtries of a nonlinear differential equation. Quaest. Math 1985;8(241).
		%
		\bibitem{mahomed1987}
		Sarlet W, Mahomed FM, Leach PGL. Symmetries of nonlinear differential equations and linearisation. J. Phys. A: Math. Gen 1987;20(277). 
		%
		\bibitem{chandru2005a}
		Chandrasekar VK, Senthilvelan M, Lakshmanan M. On the Complete Integrability and Linearization of Certain Second-Order Nonlinear Ordinary Differential Equations. Proc. R. Soc. A 2005;461(2451).
		%
		\bibitem{duarte1987}
		Duarte LGS, Duarte SES, Moreira IC. One-dimensional equations with the maximum number of symmetry generators. J. Phys. A: Math. Gen 1987;20(L701).
		%
		\bibitem{ghosh2014}
		Ghosh S, Ray DS. Li\'enard-type chemical oscillator. Eur. Phys. J. B 2014;87(65).
		
	\end{thebibliography}
	
	

\end{document}